# A CRYOGENIC LIQUID-MIRROR TELESCOPE ON THE MOON TO STUDY THE EARLY UNIVERSE

Roger Angel[1], Simon P. Worden[2], Ermanno F. Borra[3], Daniel J. Eisenstein[1], Bernard Foing[4], Paul Hickson[5], Jean-Luc Josset[6], Ki Bui Ma[7], Omar Seddiki[3], Suresh Sivanandam[1], Simon Thibault[8], Paul van Susante[9]

[1] Steward Observatory, The University of Arizona, 933 N. Cherry Avenue, Tucson, AZ USA 85751

[2] Office of the Director, NASA Ames Research Center. Moffett Field, California 9403, USA.

[3] Université Laval, Faculté des sciences et de génie, Pavillon Alexandre-Vachon, Québec Canada G1K7P4

[4] Chief Scientist, ESA/ESTEC/SCI-S, postbus 299 Noordwijk 2200 AG Noordwijk, The Netherlands

[5] Department of Physics & Astronomy, University of British Columbia, 6224 Agricultural Road Vancouver, B.C,. V6T 1Z1, Canada

[6] Space-X Space Exploration Institute, 1 rue Jaquet-Droz, CH-2007 Neuchâtel, Switzerland 2020

[7] Texas Center for Superconductivity, University of Houston, Houston TX USA 77204

[8] Immervision, 2020 University Avenue, Montreal, Québec, Canada, H3A 2A5

9 Division of Engineering, Colorado School of Mines 1610 Illinois Street, Golden, CO 80401

Shortened abstract:

*We have studied the feasibility and scientific potential of zenith observing liquid mirror telescopes having 20 to 100 m diameters located on the moon. They would carry out deep infrared surveys to study the distant universe and follow up discoveries made with the 6 m James Webb Space Telescope (JWST), with more detailed images and spectroscopic studies. They could detect objects 100 times fainter than JWST, observing the first, high-red shift stars in the early universe and their assembly into galaxies. We explored the scientific opportunities, key technologies and optimum location of such telescopes. We*

*have demonstrated critical technologies. For example, the primary mirror would necessitate a high-reflectivity liquid that does not evaporate in the lunar vacuum and remains liquid at less than 100K: We have made a crucial demonstration by successfully coating an ionic liquid that has negligible vapor pressure. We also successfully experimented with a liquid mirror spinning on a superconducting bearing, as will be needed for the cryogenic, vacuum environment of the telescope. We have investigated issues related to lunar locations, concluding that locations within a few km of a pole are ideal for deep sky cover and long integration times. We have located ridges and crater rims within 0.5° of the North Pole that are illuminated for at least some sun angles during lunar winter, providing power and temperature control. We also have identified potential problems, like lunar dust. Issues raised by our preliminary study demand additional in-depth analyses. These issues must be fully examined as part of a scientific debate we hope to start with the present article*

Abstract

	A large, zenith-pointing telescope built near the moon's north or south pole could be an extremely powerful tool to study the early universe.  The moon as a telescope site provides, in contrast to free space, a permanent, stable platform.  A fixed, zenith-pointing telescope at one of the poles would obtain extremely long exposures at the ecliptic pole, where the zodiacal background is least.  The thermal environment at the poles, with no atmosphere and the sun always at or below the horizon, is favorable for passive cooling to 100 K.  Optics at this temperature would be capable of reaching the limiting sensitivity set by zodiacal background at wavelengths up to 5 µm.  A further advantage of the moon is gravity, which opens the possibility of a primary mirror of spinning liquid with aperture much larger than that of the 6 m James Webb Space Telescope (JWST).  This would allow JWST discoveries to be followed up with detailed images and high-sensitivity spectroscopic studies. Long exposures with an aperture 20 m would yield sensitivity 30 times deeper than JWST, and 100 m aperture would be 1000 times more sensitive, allowing detection of the first stars in the early universe.

	Ionic liquids of negligible vapor pressure and very low melting temperature look the most promising to make a lunar liquid mirror.  Experiments with silver evaporated onto the liquid surface have produced progressively improved mirror surfaces.  The highest reflectivity of ~ 90% in the near infrared was obtained by preheating the liquid to form a thin skin before evaporation.  A 20 m liquid primary could be used with conventional lightweight secondary and tertiary mirror (4 m diameter) to obtain a 15 arcminute field, diffraction limited down to 1 µm wavelength.  A suspension bearing made from superconductors and magnets has been designed for very low friction in the cryogenic vacuum environment and simple experimental tests performed.  The telescope could be sited at either pole. Stars from the Large Magellanic Cloud would not cause significant difficulty for a S. Pole location near the Shackleton crater.  The North Pole also appears suitable, with ridges and crater rims within 0.5° (15 km) that are illuminated for at least some sun angles during lunar winter, and thus have the potential for solar power and heating. Lunar dust presents a potential problem that could be resolved by in-situ site testing.

## 1. INTRODUCTION

The 6 m infrared James Webb Space Telescope (JWST) and the 20-30 m class ground telescopes now being planned promise to greatly advance our understanding of the early universe. However, it will not be possible to follow up the very highly red shifted faint objects found by the JWST with any ground based telescope, because of the high atmospheric and thermal background in the infrared. This is especially true for the 1-5 µm spectral region where the zodiacal sky background is lowest and a large telescope based outside the atmosphere of the Earth could obtain extremely deep images at these wavelengths. This band has very high scientific interest for the radiation of the first stars emitted primarily at rest wavelengths from 0.1 – 0.5 and red shifted by a factor of 10 to this region.

It is possible that a cryogenic telescope much larger than JWST could be similarly orbited in free space. However, it would likely have to be constructed from components transported by multiple launches, like the space station, and the cost would likely run too high for serious consideration. The limited lifetime of around 10 years for free-flying telescopes is also a concern for such a major facility. Is there the possibility that a radically different technical approach could result in a permanent facility at radically lower cost? If we allow for the possibility of locating the telescope on the moon, where there is gravity, and can live with the constraint of observing only the sky at the zenith, then a liquid mirror telescope may be such a solution (Borra 1991, Angel 2004, Angel et al 2006).

The large paraboloidal primary mirror surface of a liquid mirror telescope is created by the natural equilibrium of a fluid spinning under gravitational and inertial forces. Very high optical quality requires simply that the vessel holding the liquid be rotated about a vertical axis at precisely uniform speed. Mechanical or thermal distortion of the containing vessel has no effect on the surface. Experience on the ground shows that such mirrors are much easier to fabricate and can be considerably lighter than their solid counterparts with rigid, prefigured segments deployed and maintained in precise alignment. Zenith-pointing telescope mirrors of liquid mercury have been made at very low cost up to 6-m diameter (Hickson et al. 1994, Potter and Mulrooney 1997). The 6-m diameter Large Zenith Telescope built at the University of British Columbia (Figure 1) was constructed for a few percents the cost of corresponding conventional telescope of a similar aperture (Hickson et al. 2007). An 8-m survey telescope is planned for Chile. The Advanced Liquid-Mirror Patrol for Astrophysics, Cosmology and Asteroids (ALPACA) will combine a liquid-mirror primary mirror with conventional secondary and tertiary optics to produce a well-corrected three-degree field of view. Equipped with a large mosaic camera, the telescope will produce drift-scan images of 1000 square degrees of sky in 5 wavelength bands to a magnitude limit of 27 to 28. The primary science goal is to study the light curves, host galaxies and standard-candle relations for various types of distant supernovae (z ~ 1) as a probe of dark energy. Other programs include gravitational lensing studies, novae, gamma ray bursts, AGN, proper motion surveys, brown dwarfs, Kuiper Belt objects and near-Earth asteroids. The project is led by Columbia University (PI: A. Crotts) with collaborating institutions including Stony Brook University, the University of Oklahoma and the University of British Columbia.

Without an independent infrastructure in place at a lunar pole, it would likely be impossible to construct and operate a large liquid mirror telescope. However, the poles are a prime candidate for future exploration of the moon, because of the presence of water and continuous solar power. It is thus not out of the question that there might be a capability to support initial construction and continuing robotic repair and instrument renewal. A unique and extraordinarily powerful telescope could give added value to a human presence endeavor, just as the Hubble Space telescope has given added meaning to the Space Shuttle program.

## 2. SCIENTIFIC JUSTIFICATION

Because a LLMT would point to the zenith near the lunar pole, it would necessarily survey only a small part of the sky, perhaps between a few and a few dozen square degrees, depending on the exact location and design. This situation is unfavorable for many science cases, such as planet characterization or the study of nearby galaxies, because the sources found in the 0.1 % of the sky available to the LLMT would generically be fainter and further away than the targets available to a steerable telescope. This means that a smaller steerable telescope, such as JWST, can achieve better physical resolution and sensitivity than a larger fixed telescope. To compensate a field of view of only 0.1 % of the sky, one would typically need a factor of 10 in linear aperture advantage. Moreover, the LLMT would not point at a ``special'' part of the Milky Way (nearby star-forming regions or the galactic center) and so is not well suited for Galactic studies. It is worth mentioning that the Large Magellanic Cloud *is* near the lunar South Pole and hence there might be applications with LMC targets; however, we have not explored this.

Fortunately, there is a major science case that avoids these limitations, namely the study of high-redshift galaxies. Put simply, there are no closer examples of galaxies at a given redshift, and so a large telescope aperture translates immediately into better spatial resolution and lower luminosity thresholds. Moreover, because of the statistical isotropy of the Universe, all reasonably sized patches of the sky are equally useful and the lunar pole is as good as any. A field size of 3 degrees corresponds to 125 $h^{-1}$ comoving Mpc at z=1, 250 $h^{-1}$ Mpc at z = 3, and 400 $h^{-1}$ Mpc at z = 15. These fields are sufficient for the study of galaxy properties across the full range of environments.

The fact that a LLMT can be cooled to around 100~K means that the telescope can be background-limited at shortward of 4 microns. The background from space has a minimum around 3 microns, nestled between the backscattering of sunlight by zodiacal dust and the thermal radiation of that dust. As highlighted by the Spitzer and JWST missions, the region between 1 and 5 microns is a superb range for the study of high-redshift galaxies (Gardner et al. 2006). Indeed, at z > 6, the scattering of light by the Lyman α resonance in intergalactic gas along the line of sight means that the objects are invisible at wavelengths shorter than 0.12 (1+z) microns (Gunn & Peterson 1965, Becker et al. 2001, Fan et al. 2001), which pushes one into near-infrared wavelengths where the sky brightness on the ground is high. Hence, for the study of galaxies at z > 15, one requires a cooled telescope to reach space-background-limited performance. Working between 1 and 5 microns allows one to isolate high-redshift galaxies through their strong Lyman α breaks using multiband photometry (for a review, see Giavalisco 2002). At lower redshifts (1 < z < 6), the 1-5 micron band permits detailed study of the rest-frame optical and indeed yields maximum photometric sensitivity for intermediate-age stellar populations.

The integration time per point depends on the LLMT design (section 4.3) and placement (section 5). The field of view of the telescope would generically be near the south pole of the moon. If displaced slightly, the field would execute a circle on the sky each month as the moon rotates (section 5.1). However, in addition, the pole of the moon precesses every 18.6 years, so the center of the circle itself tracks out a circular path with a radius of 1.5 degrees. In the limit that the LLMT field is nearly at the lunar pole, then the field moves 0.52 degrees per year. For a 20-m LLMT, the diffraction limit at 3 microns would be 0.037 arcsec, so Nyquist sampling would require about 2500 pixels per square arcsecond. With a billion pixel camera and a square field of view, this makes the field of view about 10 arcminutes on a side. Hence, the precession causes the field to drift every two months. With 6 filters, this would be an integration time of 10 days per point. One would image 0.16 square degrees per year, or 3 square degrees over the full 18-year precession. If one displaced the LLMT field slightly off the lunar pole, then one could increase the field of view at the expense of integration time. One could increase the exposure time if the imaging field was rectangular and extended along the precession axis.

Deep surveys with JWST will reach as faint as 1~nJy (31.4 mag on the AB system) in the 1 to 5 micron region with integration times of about a day per filter (Gardner et al. 2006). For comparison, a 20-meter LLMT with an integration time of 10 days per filter could reach 30 times fainter, 35 pJy or 35 mag AB, assuming a point source. A 100-meter LLMT could reach another factor of 25 fainter, 1.4~pJy or 38.5 mag AB, with the same assumptions. A 100-meter with the same Nyquist-sampled gigapixel camera would have an integration time of only 2 days per point, which would decrease the depth to 3.1~pJy and decrease the total survey field by a factor of 5. For resolved objects, the 20-meter LLMT still gains a factor of 10 in surface brightness sensitivity over deep JWST surveys, while the 100-meter LLMT gains another factor of 5 or 2.2 depending on camera assumptions.

The ability to reach 30 to 1000 times fainter than deep JWST images would be remarkable. For z > 10 galaxies, this depth translates essentially immediately into the star-formation rate being probed. Continuous star-formation of about 1 solar masses/yr at z=15 is roughly 30 mag AB at 2 microns after 10 Myr; the object gets about 0.5 mag brighter if it can sustain this star formation rate for 100 Myr (Bruzual & Charlot 2003). Hence, JWST can reach star formation rates of about 0.25 Msun/yr at z=15. This corresponds to proto-galaxies with millions of solar masses of formed stars. Since such a burst of star formation is likely only a small fraction of the baryonic composition of a galaxy and the baryons are only 1/6 of the total mass, JWST will only detect halos of about $10^8$ solar masses at best. If the star formation duration is closer to the Hubble time, then the halos would be at least $10^9$ solar masses, assuming complete conversion of the baryons to stars.

These halo mass scales are larger than an important mass scale in early galaxy formation. Halos smaller than $10^7$ to $10^8$ solar masses do not shock heat their gas sufficiently during formation to reach temperatures of $10^4$ K, which means that the gas in the halo cannot cool by collisional excitation of atomic hydrogen and helium (Haiman, Thoul, & Loeb 1996; Tegmark et al. 1997; Abel et al. 1998; Barkana & Loeb 2001). At very low or even primordial metallicity, this gas can only cool by the formation of molecular hydrogen, which is a poor coolant and is slow to form (Haiman et al. 1996; Abel et al. 1997). As a result, such halos form stars inefficiently. Only halos above $10^8$ solar masses have formation shocks with enough velocity to heat the gas to temperatures with atomic cooling. Hence, there is expected to be a transition in the properties of high-redshift galaxies around this mass scale. The extra depth of a 20-meter LLMT would allow one to probe down to 0.001 solar masses/yr star formation rates, corresponding to halos somewhat below the transition scale.

A very aggressive goal of observational cosmology would be to image the supermassive Population III stars that are theorized to form in the first collapsed halos (Bromm, Coppi, & Larson 1999; Abel, Bryan, & Norman 2000, 2002). Such stars are very hot ($10^5$ K) (Bromm, Kudritzki, & Loeb 2001), and most of the energy is released shortward of 1216 Angstrom, where it is scattered by the neutral intergalactic medium, in large part to form a low surface brightness cloud of about 10" radius (Loeb & Rybicki 1997). One can still, however, image these object in the rest-frame UV, e.g., around 1500 Angsroms. For a 1000 solar masses star radiating at the Eddington limit and $10^5$ K, one finds a luminosity density of $5\ 10^{24}$ ergs/s/Hz at 1500 Angstroms. At z=20, this corresponds to a flux density of 1.7 pJy. This may be within reach of a very large LLMT.

Another exciting possibility would be to observe supernovae from these primordial-composition stars. Wise & Abel (2005) estimate that the rate of these supernovae would be 0.1-1.5 $yr^{-1}$ $deg^{-2}$, roughly 1 per 100 JWST NIRCam fields. The predicted fluxes depend sensitively on the mass of the primordial star but could well be at or below the detection limit for long JWST exposures, particularly at z>25 . Hence, the high mapping speed of a LLMT would be enabling for this search. It should be noted that the fact that these supernovae rise and fall over years due to the cosmological time dilation means that one would need to design the LLMT to be able to revisit at least some portion of its field for years.

Alternatively, one could use a LLMT to conduct spectroscopy instead of imaging. A ten-day integration with a 20-meter LLMT on a point source could reach S/N=10 per R=3000 resolution element on a 2~nJy source, if one is not detector limited. Such spectroscopy is similar to that of the typical signal-to-noise ratio in the Sloan

Digital Sky Survey spectroscopic sample (York et al. 2000).  Hence, a 20-meter LLMT could obtain spectroscopy of high diagnostic power on the faintest objects detected by JWST.  However, achieving background-limited performance will be challenging: at R=3000, the count rate from the background is only about 3 photons per hour per resolution element per PSF.  Backing off to R=100 would ease this requirement and would allow one to push to below 1~nJy.

Another potential application of a LLMT would be to measure metal absorption lines in the intergalactic medium at $z = 10$ using higher redshift galaxies as the sources. This would allow one to probe the initial dispersal of heavy elements in the Universe and study the winds from proto-galaxies, in a manner analogous to studies at $z = 1$ and $z = 3$ today (e.g., Adelberger et al. 2003, 2005).

There is an exception to the statement that a generic patch of the sky is as good as any for the study of high-redshift galaxies.  A pointable telescope like JWST can achieve extra sensitivity by using gravitational lenses to magnify small parts of the Universe  (e.g., Egami et al. 2005; Stark et al. 2007). Magnifications of a factor of 10 can be found over regions of a few hundred square arcseconds in size around rich clusters.  In practice, this comes at significant cost: the volume being surveyed at high magnification is tiny, perhaps 10 arcseconds compared to the 10' field of view of our notional gigapixel 20-meter LLMT.  Moreover, one cannot do a filled survey and therefore one loses all control over the environment of the sources one is detecting.

Fundamentally, although JWST will be a staggering improvement (100-1000) over current observational capabilities for studying high-redshift galaxies, it will not finish the field.  LLMT technology could be a route for achieving an additional 3 to 5 order of magnitude improvement.

### 3. LIQUID MIRROR TELESCOPES

Telescope primary mirrors must achieve and maintain very tight optical tolerances, typically some tens of nm or less. Conventional telescopes achieve this, at high cost, by polishing and complex support systems. Liquid-mirrors, on the other hand, have mechanical tolerances that are three orders of magnitude lower. They achieve and maintain surfaces of optical quality simply due to the natural equilibrium of a fluid under gravitational and inertial forces. Consequently liquid mirrors are much easier to fabricate and can be considerably lighter than their rigid counterparts.

Mercury has been used to make inexpensive mirrors with excellent surface quality. The technology is young but its performance is well documented by laboratory tests (Girard & Borra 1997, Tremblay & Borra 2000) as well as by observations (Hickson & Mulrooney 1998; Cabanac, Borra, & Beauchemin 1998, Hickson and Racine 2007). Zenith-pointing telescope mirrors of liquid mercury have been made at very low cost up to 6-m diameter (Hickson et al. 1994, Potter and Mulrooney 1997).  Liquid mirror telescopes such as the 6-m diameter Large Zenith Telescope built at the University of British Columbia (Figure 1) can be constructed for a few percent the cost of corresponding conventional telescope designs of a similar aperture (Hickson et al. 2007).

The Moon provides a highly favorable environment for cold liquid mirror telescopes, as it has gravity but no atmosphere or wind to disturb the surface, and very little seismic disturbance. The lower gravity, compared to Earth, permits large light-weight structures. The lack of atmosphere also allows a new technique, not practical on Earth, in which a highly-reflective coating is evaporated directly onto the spinning liquid surface. The lower gravity also gives the minor mechanical advantage that the rotational velocity needed, for a given focal length, is lower.

Mercury is not a suitable reflective liquid for a lunar survey telescope that operates in the infrared. The reflecting surface must be much colder than the freezing point of mercury, or it will produce a background thermal radiation much brighter than the very dark natural sky background. A key requirement is thus to identify such a very cold liquid, about the temperature of liquid nitrogen, that can be made reflective with a suitable coating.

The dish of reflecting liquid we envisage to be supported by a superconducting levitation bearing. A control system would stabilize the mirror by means of electromagnetic forces. A cylindrical solar radiation shield, surrounding the telescope, would protect it from the sun and allow passive cooling to achieve the low temperatures required for operation. This would be constructed of very lightweight multi-layer insulation. The telescope would be built by some combination of astronauts and robots, taking advantage of capabilities developed as part of future lunar exploration

## 4. TECHNICAL CONSIDERATIONS

We considered a series of designs. Table 1 summarizes the design parameters. The 2-m precursor telescope would test the new technologies that shall have to be developed and will also carry out a competitive research program. The 20-m LLMT, hopefully to be followed by a gigantic 100-m telescope, would be the first competitive instrument. We consider the optical design of a three-mirror f/15-f/20 system diffraction-limited at 1 µm wavelength. The focal ratio of f/15 was chosen so the 15' field of view can be critically sampled (Nyquist sampling) at 1.6 µm wavelength with current technology arrays with 18 µm pixels. A multi-object spectroscopic instrument is also essential to meet our scientific requirement. Working in the infrared requires a very cold environment and sets stringent technical requirements on the telescope and instrumentation. For example, the liquid mirror must operate at a very low temperature (about 100K) that is above its freezing point, setting strict requirements for the liquid substrate. Also, the superconducting bearing we consider will need to be held below its critical temperature, and the focal plane will have to operate at a temperature, in the region of 30K. These temperature requirements could be met by a combination of radiative cooling and Ohmic heating. During the winter the sun, low above the horizon, may be hidden behind geographical features. In winter, power storage and/or power could be relayed from solar panels erected on high points typically 20 km distant. Issues related to the lunar location and environment are discussed in the next section.

**4.1 Reflective Liquids**

Besides high reflectivity, the liquid mirror must meet two additional major requirements. To observe below 10 microns, the thermal emissivity of the mirror must be substantially lower than the natural zodiacal sky background; therefore, the telescope ideally requires a mirror surface at < 100K. Obviously, the liquid mirror needs a substrate that remains liquid at this temperature. Finally, the liquid must have the low vapor pressure required to prevent evaporation in the lunar vacuum. We considered three main classes of liquids: eutectics of alkali metals (Borra 1992), which have the advantage of high intrinsic reflectivity, low-reflectivity liquids coated with self-assembling nanoparticles.(Borra et al. 2004) and low-reflectivity liquids coated with a thin metallic solid layer. We concluded that the most promising approach was to apply metallic coatings to low temperature, low vapor pressure liquids.

Borrowing from techniques used to coat solid mirrors on Earth, we conducted a number of experiments coating a variety of liquids by vaporizing silver on their surfaces. A recent article (Borra et al. 2007) summarizes our recent efforts. We successfully coated an ionic liquid with silver. Coating an ionic liquid was a major breakthrough because ionic liquids have negligible vapour pressures and thus do not evaporate in vacuum. We experimented with a commercially available ionic liquid that solidifies at 175 K and we are actively pursuing ionic liquids with lower freezing temperatures**.** There is a realistic expectation of success because ionic liquids are organic compounds and there are at least $10^6$ simple ionic liquids, and $10^{18}$ ternary ionic liquid systems (Borra et al. 2007), giving a phenomenally wide choice for optimizing the properties of the liquid substrate.

The reflectivities reported in Borra et al. (2007) are of the order of 80% and must be improved. We have pursued our coating efforts. Our previous experiments have shown that silver tends to diffuse in the liquid substrate, making it difficult to obtain a thick layer. Note, however, that this is a problem only during deposition and that, after deposition, the surface coating is stable. Diffusion appears to be the reason why the reflectivities of coated liquid are lower than the reflectivity of metallic silver. Previously, we found that diffusion could be reduced by first coating the ionic liquid with a thin Chromium substrate. Chromium has a high fusion temperature and was difficult to evaporate. Like chromium, aluminum is more reactive than silver while having a lower melting temperature. Consequently, it is a good candidate to replace Cr as an intermediate layer. We carried out several coating experiments but did not obtain substantial improvement.

To study the effect of heating on the liquid substrate, we carried out a series of experiments where we heated the liquid before depositing the metallic coating. The ionic liquid (ECOENG212) was first heated, in vacuum, at a temperature greater than 100 degrees C for a few hours. The liquid was subsequently coated with silver in vacuum at an ambient low pressure of the order of $3\times10^{-7}$ Torr. Our first experiments evaporated silver on the liquid maintained at a temperature above 100 degrees C but gave disappointing results. We only obtained fragile layers having negligible reflectivity. We also found that the ionic liquid evaporated if we kept it a temperature greater than 150 degrees C in a vacuum at $3 \times 10^{-2}$ Torr.

We then tried coating the liquid at room temperature. We first heated the liquid, in vacuum, at a temperature greater than 100 degrees C for a few hours but now allowed the liquid to cool to room temperature at $3\times10^{-7}$ Torr. In some of the experiments we used a layer of ionic liquid that was 3-mm thick. On these 3-mm layers (but not on thicker layers) we noticed the formation of a thin transparent skin on the surface of the liquid. The skin was detectable only because it had a few wrinkles. We then coated the skin covered liquid with silver obtaining a silver surface layer that had a high reflectivity. Presumably the thin surface skin prevents the silver from diffusing in the liquid during the coating process. Figure 2 shows a typical reflectivity curve so obtained. We can see a respectable reflectivity of 92%. However, the silver layer had a few surface wrinkles visible with the naked eye. We are now carrying out experiments to better understand the formation of the membrane and of the wrinkles.

**4.2 Bearings**

An ideal bearing should be able to support the weight of the rotating mirror with a high stiffness while offering little resistance to its spinning motion. For a rotating mirror telescope to function on the moon, it has also to work under the harsh conditions of a cold dusty vacuum. Preferably, it should also be light in weight and expend no power. With these requirements in mind, the superconductor magnet bearing is one of the most promising candidates. The superconductor magnet bearing works on the unique capability of a superconductor to keep a magnet at a fixed distance from it from the moment the superconductor and the magnet are cooled down to cryogenic temperatures in the vicinity of each other. The distance between superconductor and magnet can be of the order of tenths of an inch, which mitigates the potential problem of seizure by being clogged up with dust when deployed to the moon. A small prototype having a diameter of an inch and a weight of 0.083 daN was demonstrated at the Texas Center for Superconductivity at the University of Houston. (Angel et al. 2005) It supported the dish of liquid weighing 0.18 daN and spinning at about 50 RPM, yielding a mirror of focal ratio f/1.

A simple scale up of this simple design to support a 20 m liquid mirror telescope on the moon, where it would weigh 164 daN, would require a bearing with a mass of 2400 kg. The bearing mass of this simple design could be reduced to 300 kg by using a larger diameter superconductor disk, but reducing its thickness, with similar reduction of the magnet dimensions.

Simplifying from an earlier work by Lamb et. al., we have designed a bearing with a mass of 25 kg by making use of a supplementary magnet to take over the weight bearing function off the high temperature superconductors. This pared down both the amount of superconductor and total magnet materials required, as the force density between magnets are much higher than that between magnets and superconductors in this application. We cannot do away with superconductors altogether. They are still needed for stabilization of the force between the magnets.

A schematic diagram of the design is shown in Fig. 3. The three main components are: (1) a stator magnet, fixed with respect to ground, (2) a rotor magnet, to which the rotating liquid mirror telescope is to be attached, and (3) high temperature superconductor (Lamb et al. 1995) stabilizer where indicated in Fig. 3. The magnets are rare-earth permanent magnets, made of an alloy of neodymium, iron and boron. They should carry a magnetization of 0.8 MA-m in the axial direction. The HTS stabilizer is made of the ceramic, yttrium barium copper oxide (YBCO). They are made in the form of rings fitting into the space as shown in Fig. 3.

Three more pieces of accessory equipment are needed, but not shown in Fig. 3: (1) a drive to rotate the liquid mirror telescope without direct physical contact, (2) a hold and release mechanism with a backup bearing, to hold the rotor magnet in place before activating the bearing during assembly, and (3) some means of cycling the temperature of the embedded superconductors in the range 77 – 100°K.

In the deployment of the rotating liquid mirror telescope, the superconductor magnet bearing should be the last component to come on line before the rotation of the liquid mirror. To start with, the superconductors should be warm, at or above 100°K. The rotor magnet, with the telescope attached and liquid filled, is held in the position inside the stator magnet as shown in Fig. 3 with the release and hold mechanism. If the rotor magnet is released, the telescope should just hang, but it would also cling to the side of the cavity of the stator magnet rather than staying centered. With the rotor magnet held to the center, the superconductors are then cooled to 77°K. It is now safe to release the rotor magnet. The flux pinning force between the rotor magnet and the cold superconductors will maintain the centered position of the rotor magnet. Then, the liquid mirror telescope is ready to be spun up to speed.

In our previous work, (Angel et al. 2005) we have shown that even when the rotation of the mirror approaches a period equal to that of the pendulum period of the suspended mirror, no instability of the surface of the rotating liquid ensues despite resonance of the rotational motion with the pendulum swing. The radius of curvature of the liquid surface was then equal to the pendulum length with the fortunate consequence that the wobble of the mirror at the pendulum period did not upset the natural curvature of the surface, which remains fixed with respect to the dish, a natural way to achieve a stable mirror.

Once up and running, no maintenance is required. No inherent failure modes are known, though this must be qualified with the limited history of the superconductor magnet bearing. The longest that we have had one in continuous operation is six days and nights.

To achieve the much more severe optical requirements for a diffraction-limited telescope, such as the spinning axis remaining vertical to within ~ 1 milliarcseconds, the gap stable to within a micron, and to avoid defocusing, a required inherent axial stiffness of ~ 40 N/mm and a tilt compliance ~0.02 sec of arc/mN*m, we would still have to resort to active control. With the use of the superconductor magnet bearing, we would have a more stable platform upon which to build the required active control that would be less demanding on power requirements.

**4.3 Optical Design**

One of the main challenges of the optical corrector for the lunar telescope is that it must be diffraction limited in order to fully exploit the surrounding environment. We have first explored off-axis systems that would allow increased sky cover and minimize light loss due to obscuration. These however generally present large field distortion, typically at the level of a percent or so. But once it exceeds the reciprocal of ten times the number of resolution elements across the field ($\sim 10^4$), it results in unacceptable image degradation during long exposures. This precludes drift-scanning to accommodate star motion away from the zenith, or any kind of long exposures except for the special case when the telescope is located *exactly* at the Lunar pole or otherwise set to image only a set zenith position, so that a rotation about the symmetry axis of the primary mirror can be made to compensate for Lunar rotation. Designs to circumvent this problem exist but require complex off-axis optics and sophisticated control systems. We conclude at that time that *off-axis LLMT's are not practical*.

A large LLMT should therefore have simple optics that view the zenith. However the prospect of accurately deploying, or assembling, a large complex corrector at the top of a ~ 30 m high structure on the Moon is daunting. For this reason a Cassegrain type system, with a small single mirror deployed above the primary, would be preferred. An example of one such design, with three reflecting mirrors is shown in figure 4.

This three mirror design achieves the goal of a 15-arcminute field with diffraction limited images at $\lambda \geq 1$ micron. The design has an f/1.5 20 m parabolic liquid primary, a 2.4 m Cassegrain convex secondary is located 26 m above the primary and a 4 m concave tertiary mirror. The final focus is at f/15, for a field diameter of 1.3 m. The imaging array would be a mosaic of 36 x 36 four-megapixel arrays, with each 18 micron pixel corresponding to 12.4 milliarcseconds. This provides Nyquist sampling of images down to 1.6 µm wavelength. This optical concept has significant vignetting by the focal plane, and options to reduce this by managing the pupil position of the LLMT are being explored or by tilting the tertiary mirror. This type of design is also expandable to larger sizes – perhaps as large as 100m diameter. The distortion produces by the LLMT design for the 15 arc minute field is presented in figure 5 exaggerated by a factor of 100.

For a telescope located precisely at the pole, rotation of the star field about the optical system axis would be exactly compensated for long exposures by rotation of the detector. However, if the telescope were only 7 arc minutes (3.5 km) from the pole, the star field and the detector would rotate about one edge of the telescope field. The smear caused by distortion for the above design would be about 1 pixel/hour, about the limit we could accept.

# 5. LUNAR LOCATION AND ENVIRONMENT

We assume that the observatory will be located near one of the lunar poles, which are ideal locations for two main reasons. Firstly, they will allow relatively large fields of view and long integration times because the 18.6 year lunar precession will slowly move the field of view around the ecliptic pole in a circular path of 3º diameter, at a rate of 0.5º per year. Secondly, because the fields observed by a lunar zenith telescope are located near the ecliptic poles. The ecliptic poles are regions where the infrared sky background from zodiacal dust is minimized, rendering them ideal locations for deep infrared observations.

## 5.1 Sky Coverage

We consider a zenith pointing telescope having a corrector capable of obtaining diffraction-limited images within a 15 arc minute field. The field of view and integration times depend on the exact location of the observatory. A zenith telescope located at a pole will observe a field centered 1.55 degree off the ecliptic pole that rotates around its center once per month. Over the course of the 18.6 year lunar precession cycle, it will trace a circle around the ecliptic pole. The field observed consists of an annulus, centered on the ecliptic pole, having a circumference of 9.7 degrees and a width of 15 arcminutes. The integration time is about 6 months. For an observatory located at a small distance from the pole, the zenith sweeps around in a circle once per month, while the center of the circle slowly moves around the ecliptic pole. After the 18.6 years lunar precession, it will observe a total of 2 square degrees with integration times of about 6 months A location of especial interest would be such that the zenith is located at a latitude of 1.55 degrees from the lunar pole. The telescope will observe a 9.7 degree annulus once each lunar month centered on the lunar pole position, but passing through the ecliptic pole monthly for about 1 day a month. Over 18 years this would provide a 6 month total integration on a 15' square image of the ecliptic pole, dropping off the typically 3 weeks over a 30 square degree area. Fig. 6 shows the sky coverage.

## 5.2 Contamination by Stars in the Large Magellanic Cloud

Our initial inclination is to place the LLMT at the South Pole because it contains peaks of eternal light along the Shackleton crater, necessary for powering the telescope, and can be thermally shielded effectively by placing the telescope inside the permanently shadowed portion of the crater. Moreover, Shakleton crater has been preliminarily chosen as a lunar outpost location by the US Vision for Space Exploration. However, this location offers a view of the sky that also includes Large Magellanic Cloud (LMC), which could negatively impact our scientific case because we may be confusion-limited. Even though the North Pole does not suffer from this impediment, we do not yet know if it has any peaks of eternal light, though as discussed in Section 5.4, our first attempts in uncovering these eternally lit sites have been promising. To assess the scientific utility of the South Pole site, we quantify the degree to which the LMC contaminates our field.

Because the primary mode of operation of the LLMT is to observe a relatively small field for long periods of time in order to find faint and distant objects, a dense, nearby stellar population in that field can render much of that field unusable. A cursory look at the Digital Sky Survey (DSS) R-band plates of the North and South Ecliptic Poles shown in Fig. 7 makes this issue clear. At the DSS R-band limiting magnitude of 21, we see significant stellar contamination from the LMC in the southern field, whereas the northern field is relatively clear as it looks away from the galactic plane. However, the confusion limit is a strong function of limiting magnitude and angular resolution. The DSS plates only have a 2" resolution while the LLMT is designed to have 0.05" resolution and will reach a 26 magnitude limit at 5µm. At this sensitivity, we will be able to detect all of the main-sequence (MS) stars in the LMC and go a magnitude fainter. Even though we will detect more stars than DSS, it is unclear if we will be confusion-limited because our angular resolution is a factor of 40 better.

To address this issue quantitatively, we model the LMC stellar population and estimate its stellar surface density to place constraints on the size of the usable field. We use STARFISH (Harris et al. 2001), a code that generates a stellar population and its associated photometric quantities from a given star formation history (SFH), to model the LMC stellar population. Zaritsky et al. (2004), who have generated an extensive stellar catalogue of the LMC, have shown that the built-in SFH for the LMC adequately reproduces the observed bright-end stellar distribution. We take this model and extend it to the faint end of the MS by assuming a Salpeter IMF and a lower mass cut-off of 0.1 $M_\odot$, as the LLMT will be able to detect and resolve all of the stars in the LMC, to get a crude measure of the stellar densities we would observe.

We take a 5' patch centered on the Southern Ecliptic Pole (SEP) and count the number of stars found in the Zartisky et al. (2004) catalogue, which has a V-band limiting magnitude of 22. We derive a stellar density of 0.005 stars/arcsec$^2$, and when we scale this quantity to include all of the MS stars the stellar density increases to 0.3 stars/arcsec$^2$. At M-band resolution, this corresponds to a 99.6% usable field. The LMC stars only make a minimal impact on the observed field primarily because the southern field looks through a low density region of the galaxy.

We test our conclusion by analyzing an archival HST ACS High Resolution Camera image, shown in Fig. 8, taken near the SEP with the F606W filter. This HST exposure can detect V=26 at 4 sigma and consequently is deeper than the original DSS plate. Unlike the DSS exposure, the HST image, which has a resolution comparable to the LLMT, clearly shows that all of the stars are resolved. We further test our calculations by computing the stellar density inferred from the HST image and extrapolating that to the faint-end limit, which yields a result of 0.7 stars/arcsec$^2$. This roughly corresponds to our original value within a factor of 2, adding credence to our initial calculation. In this case, 99% of the field is useable. The main sources of uncertainty in this calculation arise from the imprecise determination of the LMC SFH and low mass, faint-end stellar/brown dwarf distribution. However to obscure the field-of-view to a point where it is 75% useable, we require a factor of 20 more sources. Such a large discrepancy in our modeling is unlikely. We conclude stellar contamination by the LMC is not a major issue.

One of the main science drivers for this telescope is to perform precision cosmology, which requires precise photometry (at the 1% level) of objects such as distant supernovae. Photometry at this level may be difficult to achieve if the extinction from the LMC's interstellar medium is not accounted for. A recent Spitzer survey, SAGE, carried out a large imaging survey that mapped the entire LMC in the short, mid, and far-IR partly in an effort to characterize the dust properties of the LMC interstellar medium to very high sensitivies ($A_V > 0.2$) (Meixner et al 2006). In their released data, we find that the LLMT field lies far enough away from regions dominated by both warm and cold dust that we can make accurate predictions of the dust extinction, which should be relatively low, along a given line of sight from the Spitzer data and achieve the necessary accuracy for supernovae photometry. Therefore, both poles are suitable for the primary scientific goals. Secondary science objectives and other pragmatic considerations are likely to influence the choice of location.

### 5.3  Lunar Dust

Any form of dust atmosphere on the Moon will impact the LLMT's quality of observations. Low-level dust can coat the optics, reduce throughput, and damage components. High-level dust will elevate the infrared sky background and reduce the projected sensitivity of this telescope, undermining its scientific goals. Several studies have presented circumstantial evidence for both types of dust atmospheres (e.g., Rennilson and Criswell 1973, McCoy and Criswell 1974, McCoy et al. 1975, and Severny et al. 1975). All of these studies are based on only a handful of observations where the presence of dust was not directly detected. Nevertheless, we present the current case for a lunar dust atmosphere and the purported mechanisms that keep it in place.

The primary mechanism for replenishing dust invoked by the above studies involves electrostatic levitation. Photoelectric charging of the lunar surface by high-energy solar flux levitates poorly conducting, fine-particulate dust, the main constituent of the lunar regolith. Potential differences along the surface due to variations in the incident solar flux, particularly along the terminator, promote the movement of the charged dust. Rennilson and Criswell (1973) posit this mechanism to be the source of the horizon glow observed by the Lunar Surveyor 7 Lander. The glow, observed a few hours after sunset, is considered to be forward-scattered light from a dust layer levitated 10-30 cm vertically, consisting of 5-6 micron particles. Pelizzari and Criswell (1978) develop a robust model for dust transport that is a strong function of solar illumination, which can only explain low-level dust transport. However, the solar flux is greatly reduced due to the high obliquity of the Sun's rays at the poles, the proposed location for the LLMT, suggesting that this form of dust transport may be insignificant there.

The more serious problem is the existence of an extended, high scale-height dust atmosphere, which may dash any competitive advantage of the Moon being a dark site should it prove to be true. Apollo 15, Apollo 17, and Lunokhod-2 (Severny et al. 1975) observations do suggest an extended dust atmosphere. In the case of Apollo 15, the astronauts observed an excess of light while observing the solar corona after orbital sunset. McCoy et al. (1975) interpret this as evidence for a dust atmosphere with a scale-height of a few kilometers. Apollo 17 astronauts observed streamers seemingly originating from lunar features near the terminator as they approached orbital sunrise. McCoy and Criswell (1974) again consider this as evidence for an extended dust atmosphere but with a larger scale-height of at least 100 km. Stubbs et al. (2005) have recently modeled this observation and confirmed a scale-height of 100 km is physically possible. However, an Apollo 17 astronaut disagrees with this interpretation because he considers this phenomenon to be related to the Sun where the Moon had only acted as an occulting disk (Schmidt, private communication). A more convincing detection is the in-situ measurement of sky brightness by Lunokhod-2, a Soviet lunar lander. Severny et al. (1975) show that Lunokhod-2 measured an unusually high sky background that depended on the zenith angle of the Sun, which is a characteristic signature for a scattering atmosphere.

**5.4 Solar Power and Other Location Considerations**

The choice of the LLMT site is largely influenced by our scientific goals and pragmatic constraints. Scientifically we are driven to situate the telescope at or close to the poles, but in addition to that we require the telescope to have a consistent power supply throughout the lunar day-night cycle, be easily accessible both physically and via communication links, have low ambient temperatures to cool the optics, and be isolated from any surface activity, which may kick up dust and cause vibrations. We consider the possibility of placing the telescope at the South Pole within the permanently shaded region of the Shackleton crater (Bussey et al. 1999), but find it not to be ideal because it is permanently out of the Earth's line of sight - hindering communication, requires beamed power, is not easily accessible, and is the prime location for future volatile extraction missions.

The best alternatives are polar locations, known as "peaks of eternal light," with nearly continuous solar illumination throughout the lunar day-night cycle and seasons. Based on our earlier conclusion that the North lunar pole offers a deep view of the universe with fewer foreground stars, we focus on finding suitable sites near this pole. Using 1994 Clementine data collected during winter in the South Pole and summer in the north, Bussey et al. (1999, 2005) performed a detailed study of lunar polar illumination. Bussey et al. (2005) identify permanently illuminated regions, shown in Fig. 9, during summer at the North Pole. The pole is marked at the end of a 3 km long east-west ridge where the color red signifies 75 to 90% illumination throughout the summer months. There are several promising sites on the ridge itself and the adjacent crater rims. However, Clementine did not continue operating during lunar winter in the North and therefore there is no illumination information for that period.

In a collaborative effort with the ESA's SMART-1 lunar orbiter team, we obtained the first images taken of the North Pole in winter to study the suitability of sites. The SMART-1 probe arrived at the Moon in November 2004 during the middle of winter in the North Pole. It has since spiraled down to a polar orbit and impacted the moon. We analyzed images taken by the SMART-1 $5° \times 5°$, visible-light AMIE camera during the commissioning period of the mission. We paid particular attention to images taken close in time to the January 25, 2005 winter solstice. We find ridges and crater rims within 0.5° of the North Pole that are illuminated for at least a few sun angles during lunar winter. As an example, we show an image of the North Pole taken on January 19, 2005 from an altitude of 5500 km in Fig.10. In fact the illuminated crater rim close to the Pole shown in this image corresponds to one of the permanently sunlit areas in the 1994 Clementine summer data (Foing et al. 2003). This result suggests that solar power can be available for the LLMT from locations within 15-20 km of the true pole, which can be delivered to a polar site either through cables or microwave links.

One must also consider the effects of the thermal fluctuations in the environment of the observatory. If, as we suggest, it is located near the poles, it is subjected to a maximum sun elevation of about 1.5 degrees. Consequently, we do not expect that the ground under the telescope shield will undergo significant monthly or annual swings. Consider also that, at a polar location, the ground never received much solar energy and, after the telescope is installed, it will never see any and will cool off like the bottoms of polar craters to about 50 degrees Kelvin. Conduction through the lunar soil is likely to be low. We therefore estimate that the thermal stabilization problem is not unlike that for JWST, where the sunshield protects the cryogenic telescope from the direct heating by the sun.

**5.5 Precursor Missions**

While presently available data suggest that the North lunar pole offers superior benefits, we must determine precise candidate locations. Existing Clementine data and ESA SMART-1 data should help to determine a range of potential sites. A low cost lunar Lander will be needed to evaluate in situ the most promising candidate. Millennium Space of El Segundo, CA has designed lunar Lander missions with estimated costs of $50-100M US (Fig. 11). We envisage a payload mass of no more than 10 kg. Similar low-cost missions are being developed by NASA (Marshall, 2007) and a site survey of the type we propose could be an early demonstration of this approach.

The most important task of the precursor will consist in quantitatively assessing the overriding issue of naturally levitated dust. It will determine whether the sky background is raised by a dust atmosphere by observing the sky background in the infrared and visible bands through both lunar winter and summer. It will also determine whether ambient dust presents a serious threat to the optical elements, including the liquid mirror. At the end of the mission, it would simulate human and robotic activities to determine what level of activity is acceptable before dust becomes a problem.

The precursor will evaluate line of sight views of earth, needed for communication purposes. Data transfer is a concern because of the moon's inclined orbit, the Earth is visible from the poles for 2 weeks every month. Possible solutions are to relay the data through lunar polar orbiting relay satellites or to a high point some 170 km on the near side where there will be permanent line of sight to the Earth

The telescope will inevitably suffer some periods out of direct sunlight when the winter sun passes behind elevated features on the horizon. It will therefore require power storage and/or power being relayed from solar panels erected on one or more high points. Because solar energy is needed for temperature control the precursor will determine the horizon location, and possible obstacles relative to the sun.

Assuming successful site survey missions an intermediate step could be to construct and deploy automatically on a lunar pole a 2m class LLMT. It would carry out a scientifically useful deep survey and validate the engineering studies prior to deployment of larger LLMTs. We have produced preliminary designs for a 1.7 meter wide field telescope having a diffraction limited resolution of 0.3 arcsec at 2.5 microns. It has a 3 degree annular field, 14 minute wide which would be fully covered by a ring of forty 4096 square detector arrays with 0.2 arcsec pixels. Like the present Sloan digital sky survey camera, the different detectors would have different filters and provide up to 40 wavelength bands. In any one band it would observe about 25 times deeper than the Spitzer Telescope (e. 20 nJy at 3.5 µm). We have carried out a preliminary study that suggests a relatively low weight and limited complexity.

**CONCLUSION AND DISCUSSION**

We have carried out a study of the scientific potential and feasibility of zenith-observing infrared Lunar Liquid Mirror Telescopes (LLMTs). Our conclusions are that infrared LLMTs located at one of the lunar poles and having diameters between 20 and 100 meters may be feasible. Such gigantic telescopes could obtain unique observations of the early universe, providing important data for fundamental astrophysical and cosmological research. We have also identified a number of important issues that shall have to be addressed.

Zenith Lunar Liquid Mirror telescopes present advantages and disadvantages. Their principal appeal comes from their simplicity, a major asset for a telescope that must be shipped to, and assembled on, the moon. Simplicity comes from the fundamental fact that liquid surfaces are smooth, follow equipotentials and that the basic inertial forces of Nature (gravity and centrifugal) conspire to give the liquid surface the right shape. Small perturbations are quickly corrected by liquid flow. Shipping and assembly will also be simple. Consider for example that the uncoated mirror can be shipped inside a drum to be poured on the rotating container and then coated. The rotating container holding the liquid could be an umbrella-like robotically deployed structure, a feat facilitated by the fact that the surface of the container needs to be accurate only to a fraction of the depth of the liquid. For a realistic liquid thickness of one millimeter, it would only have to be assembled and kept parabolic within a few tenths of a millimeter. Simplicity will also facilitate maintenance. Mass is a major cost driver for space expeditions. Assuming a density of 1 and a 1-mm thickness, the respective masses of a 20-m and of a 100-m mirrors would, only be 314 kg and 7900 kg. We assume advanced materials, such as graphite fiber epoxy composites, that are both strong and light weight.

Although the technology is young, it is well-established both in the laboratory and in observatory settings, where LMTs having diameters as large as 6-m have been built and operated. On Earth, their low cost (two orders of magnitude less than comparable glass mirrors), simplicity and ease of operation are well-established. Besides low-cost, the simplicity of a LLMT also gives a timeliness advantage for it could be built well before a solid mirror free-flyer or a tiltable solid mirror telescope on the Moon.

On the other hand, Zenith LLMTs also have disadvantages and several issues must be solved before building one. Their main limitation comes from their small field of regard. Since a polar zenith LLMT only observes within a limited annulus near the lunar pole, it is not suitable for general purpose observations of the type possible with free-flying observatories or lunar steerable telescopes. Although this limitation is not as important for some fields of Astronomy, Cosmology being an example, there should be debate among the scientific community to ascertain whether, in a world of limited funds, the field of regard limitation is worth the cost of a LLMT. Note however that technological improvements may increase greatly the field of regard of the next generation of LMTs, as discussed in the supplementary information section of Borra et al. (2007).

Lunar dust remains the most significant potential show-stopper. This includes high-altitude dust that may produce a strong infrared thermal background that would eliminate any advantage a LLMT might have for investigating the early universe, as well as surface dust that may coat the optics.  The lunar robotic Lander mission (section 5.5) could determine whether these problems exist. Interference from other human and robotic activities on the moon is a minor concern. Our experience with mercury mirrors indicates that surface ripples caused by mechanical disturbances are a minor problem, easily controlled by using thin liquid films and/or viscous liquids.  If dust produced by human activities turns out to be an issue, one could designate a region near one lunar pole a scientific reservation free from large scale human activities. Note that even if dust turns out to be a problem for an infrared telescope, one may still want to consider a telescope optimized for the visible or ultraviolet (Borra 1991).

There remain numerous engineering issues in mechanical and thermal design and robotic and human construction that need to be better understood. The feasibility of the program and acceptable cost ultimately rests on robotic technology.  The fully robotic assembly of a system as complex as the LLMT is beyond the present state of the art of robotics. Tele-robotic assembly is easier to carry out when humans are relatively nearby the structure to be built. Tele-robotic operations from earth will be difficult because of light-travel delays. This would be minimized by having human operators based on the moon or in lunar orbit, with communication by lunar relay satellites.

We need to refine potential lunar LLMT site options from current data, and understand what new remote observations or in situ site survey instrumentation and procedures are needed to qualify the best site.

Finally, there may be other significant technical limitations with our LLMT concept that we have not identified in this work. Precursor missions should identify them.

In a world of limited funds and resources LLMTs must be shown to be cost effective competitive instruments to conduct unique frontier research. Only then would they merit high priority in future scientific community recommendations such as decadal National Academy reports. The full demonstration of scientific potential is of critical importance. One must prove that the LLMT is a large scientific step beyond JWST. We can use the work that was done in preparing the science case for JWST as a baseline for assessing the benefits of a large lunar telescope. We will also need to model the telescope performance and the expected images of the distant universe, as has been done for JWST. This could be done by workshops attended by leading researchers in early universe Astronomy as well as instrumentation and lunar science experts. To fully appreciate the LLMT, astronomers must face the fact that they are used to tiltable telescopes and have a natural bias towards observing programs that use them. This bias should be overcome to deal with fixed zenith telescopes. Supernova surveys are an example of scientific research done with tiltable telescopes that could easily be done with a zenith telescope (Borra 2003, Corasiniti et al. 2006). Both articles give other examples of research that can be done with a zenith telescope.

Despite the aforementioned issues and limitations we believe that a LLMT should be a high priority possibility for future lunar activities. Our preliminary study suggests that the potential scientific payoff is significant and appealing.

## ACKNOWLEDGEMENTS

The authors would like to thank the NASA Institute for Advanced Concepts and the Canadian Space Agency for supporting the initial concept work for a LLMT.

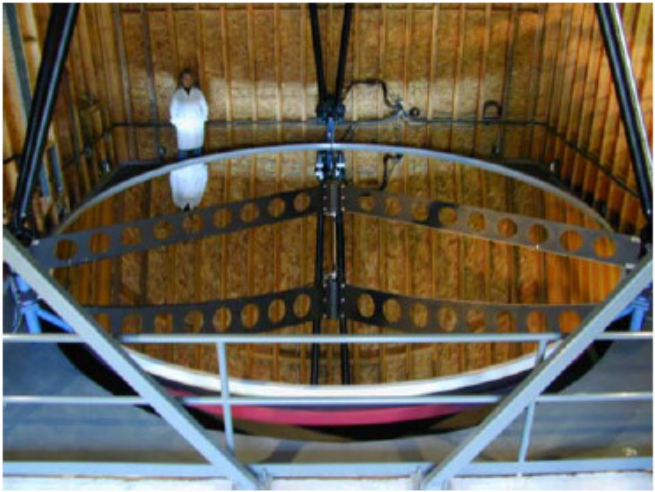 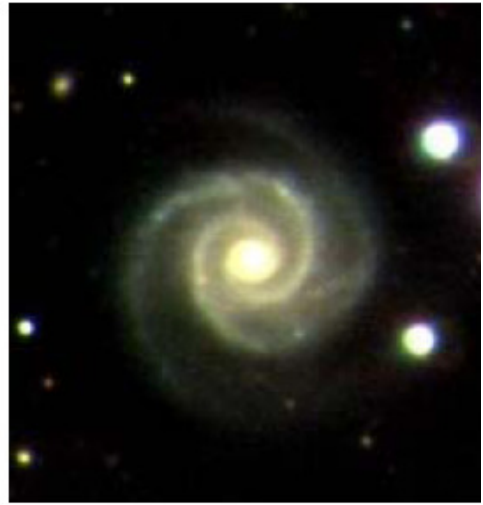

**Fig. 1.** ( (a) The 6 m diameter liquid mirror telescope at The University of British Columbia, (b) Galaxy image from this telescope.

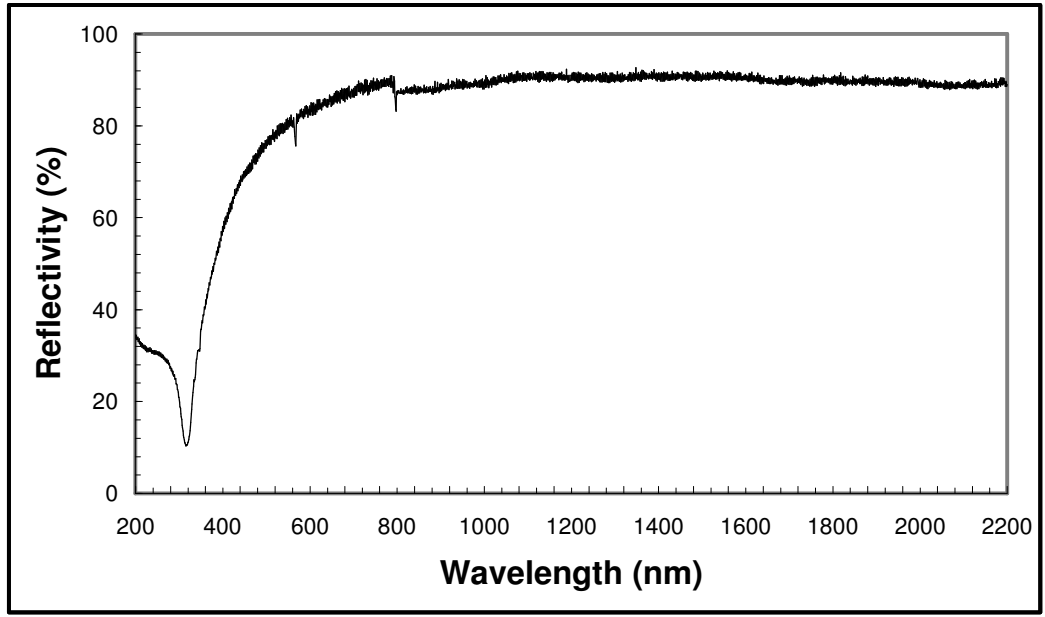

**Fig.. 2.** It shows a reflectivity curve obtained for a silver-coated ionic liquid. The curve does not extend beyond 2.2 microns because of the sensitivity limit of our spectrophotometer. We assume that the reflectivity increases monotonically to longer wavelengths.

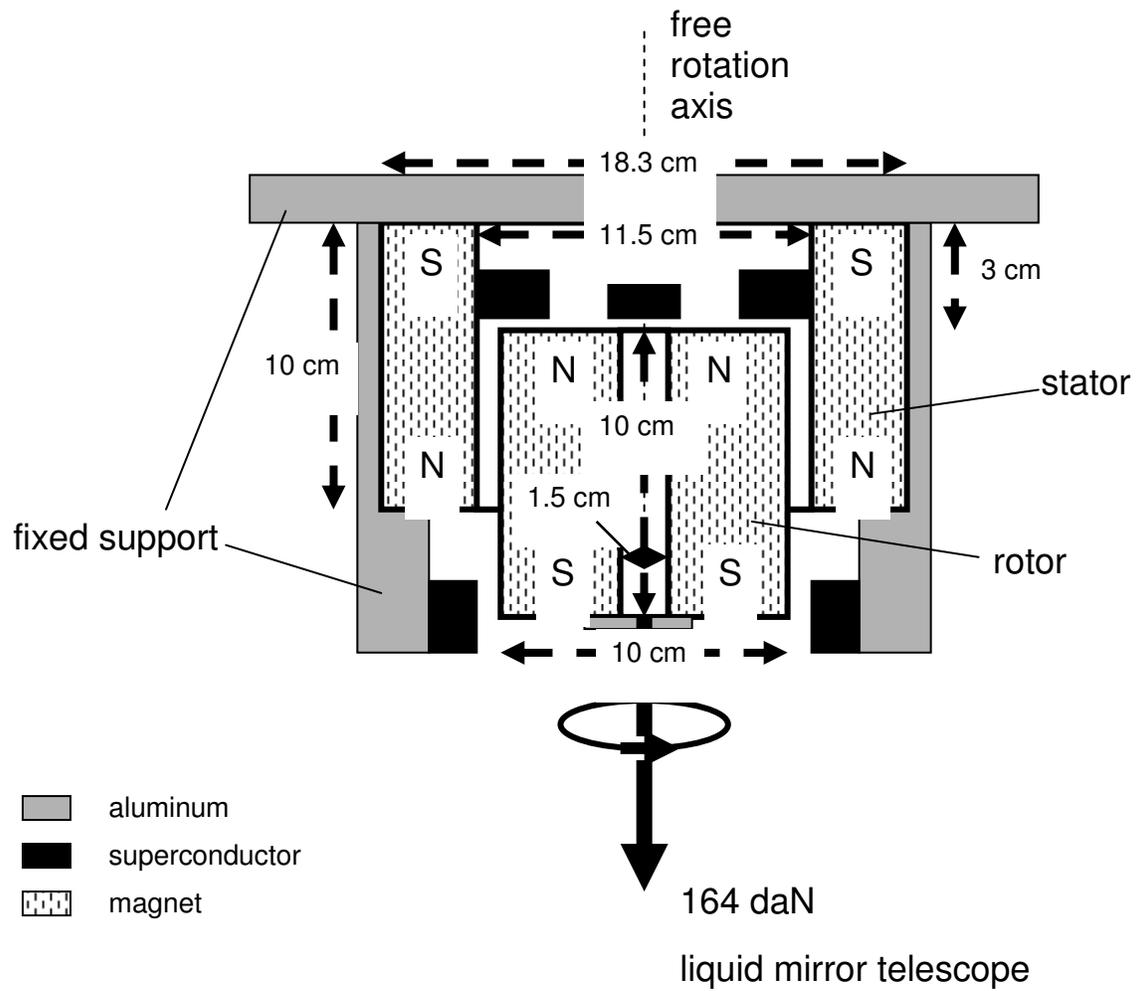

**Fig. 3.** Schematic drawing of superconductor magnet bearing designed to support a liquid mirror telescope weighing 167 kgF.

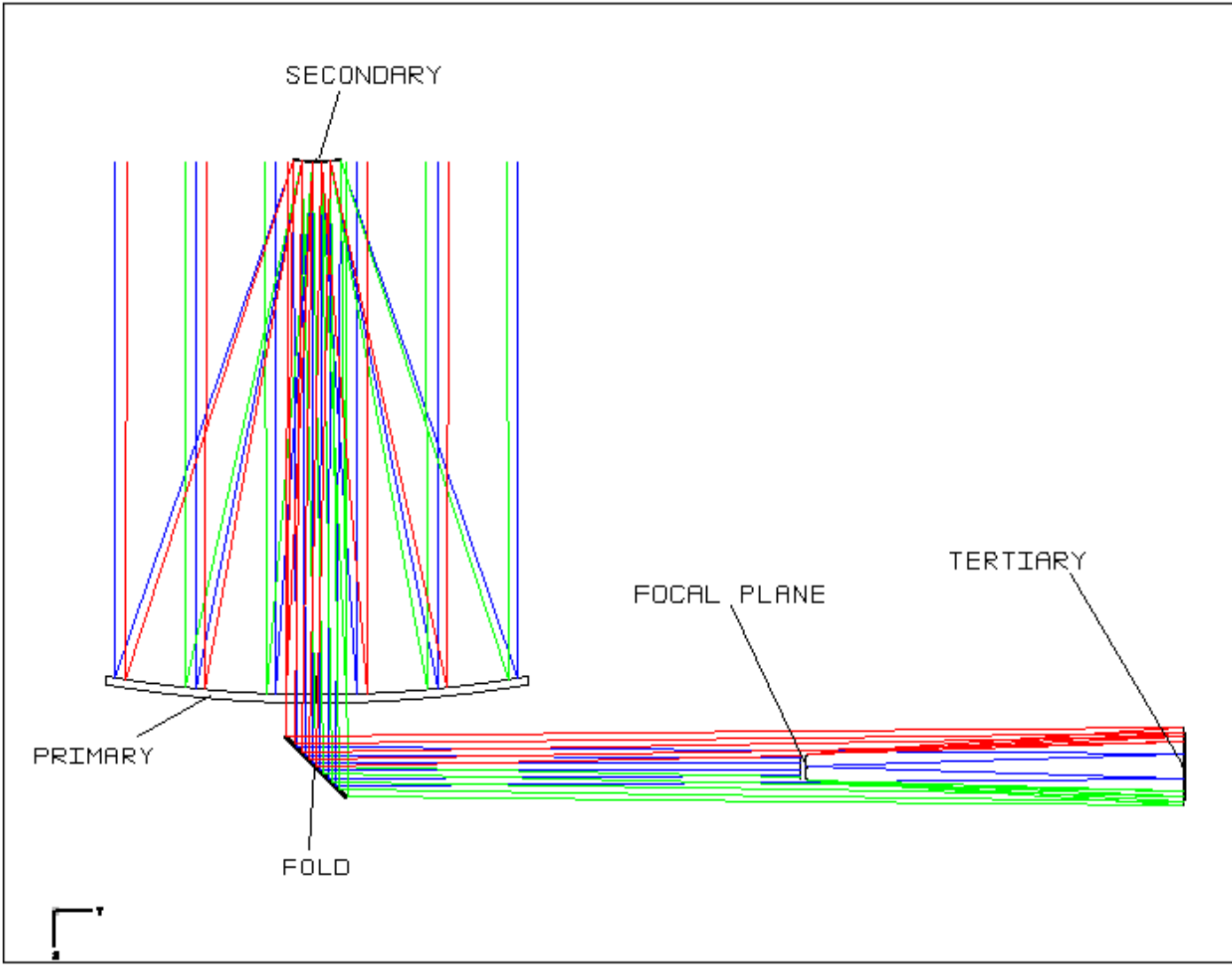

**Fig. 4**. Three-mirror telescope design

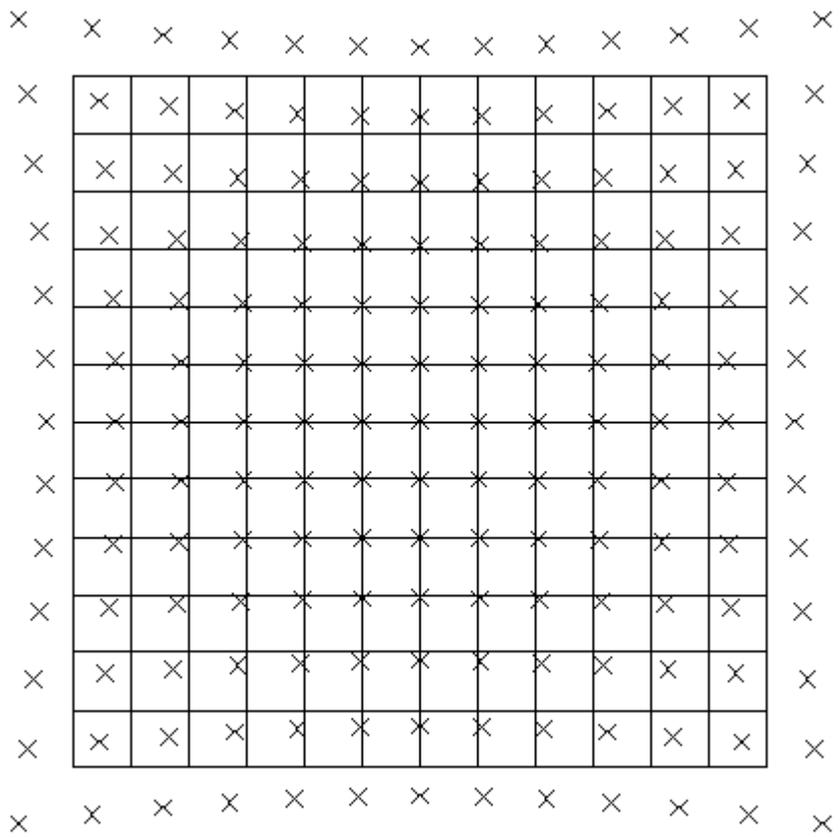

**Fig. 5.** Image distortion for the 3-mirror design shown in Figure 4.

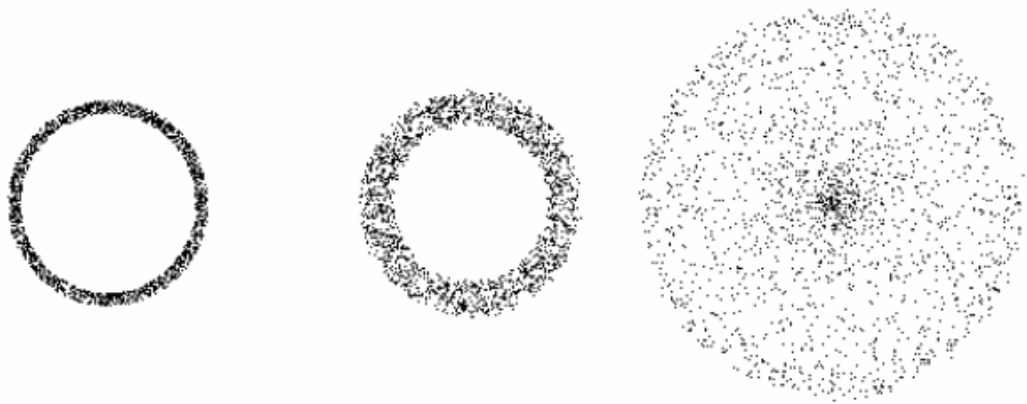

**Fig. 6.** Effect of location on sky access for a zenith-pointed telescope with 0.2 degree field of view. The density of dots is proportional to the integration times. *Left* – at the pole the field sweeps out an annulus 3.1 degrees in diameter centered on the ecliptic pole. *Center* – 0.2 degrees from the pole, the field sweeps out a half degree annulus each month. *Right* – 1.55 degrees from the pole. See text for more details

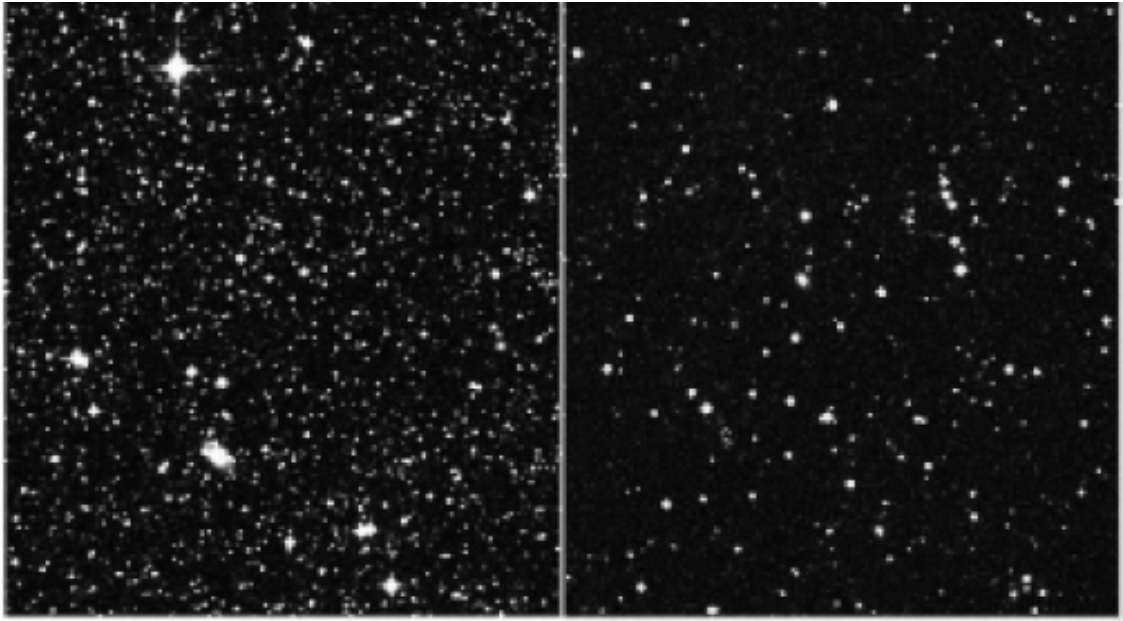

**Fig. 7.** Comparison of N and S ecliptic poles. 12' by 12' Schmidt survey photographic R-band images; Left - South Pole view, Right - North Pole view. The R-band limiting magnitude is 21 and the resolution is 2". It is clear there is significant stellar contamination in the southern field.

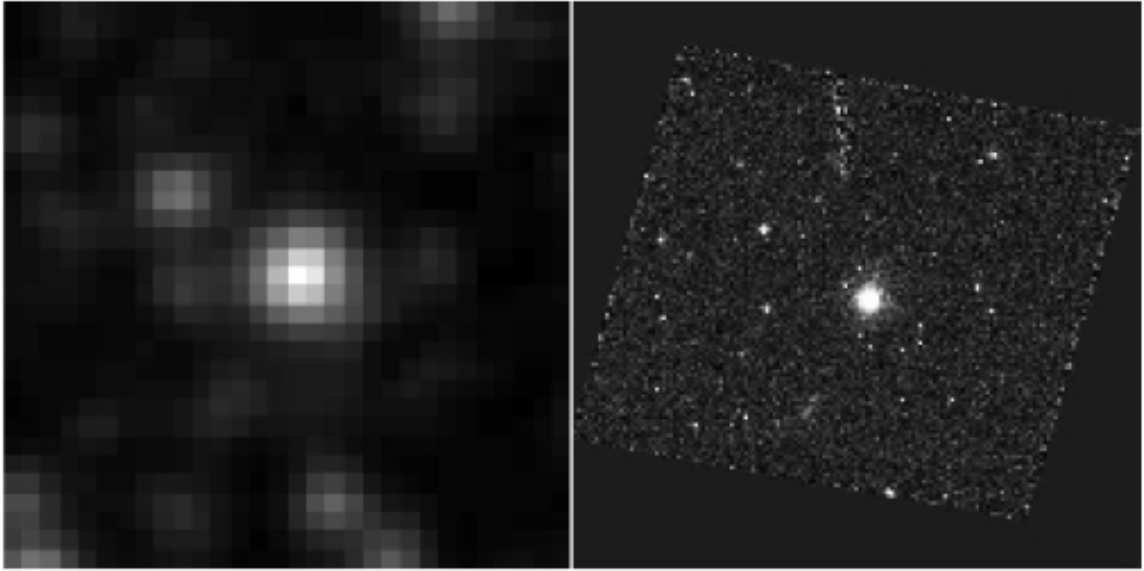

**Fig. 8.** Left: Detail from the red Schmidt survey plate of Figure 7 shows a field (30" x 20" in size) close to the South ecliptic pole. Right HST red image of the same field at the same scale – faintest stars V=26. We see that all of the stars in the HST image are resolved and are not confused.

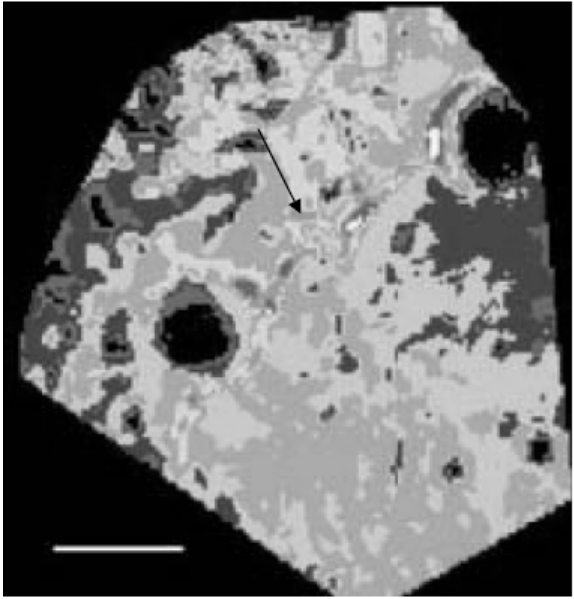

**Fig. 9.** Summertime illumination map generated from Clementine data of the North Pole (marked by arrow). The scale bar on the left is 15 km in length. The crater located in the upper right of the pole is the illuminated crater shown in Figure 10.

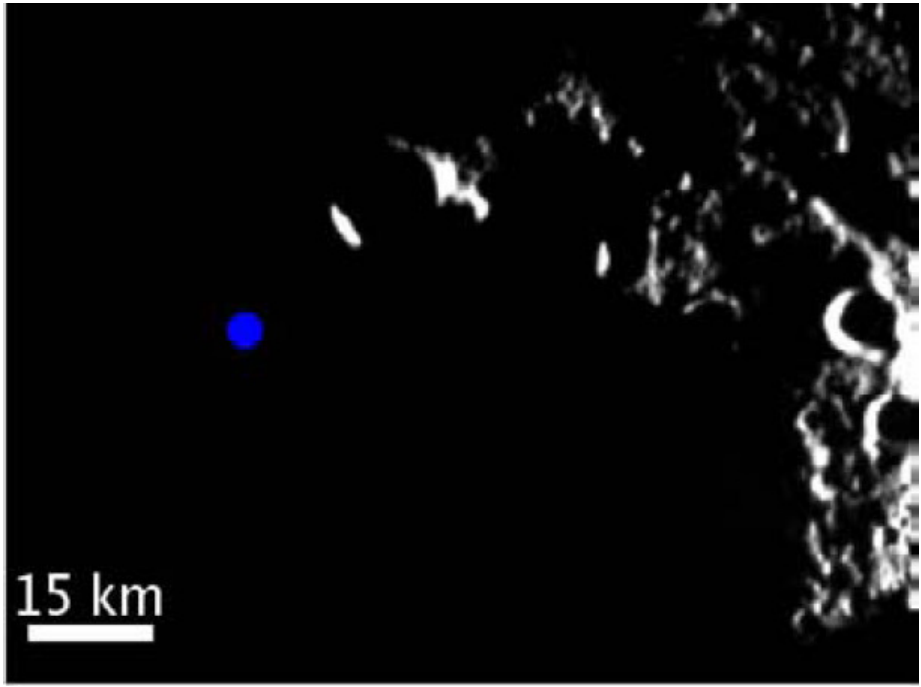

**Fig. 10.** SMART-1 AMIE image of the North pole in mid-winter. This is the average illumination map obtained in summer, from permanently shadowed areas (black) to 99% (white spots near the pole and crater rim). The dark pole is at the center-left, and the nearest crater is still showing illumination. 15 km distant is the top-right crater of Figure 9, which has 100% summer illumination. Solar panels on this ridge would provide power for at least some of the winter months (Credit: ESA/SMART-1/AMIE team/Space-X Space Exploration Institute).

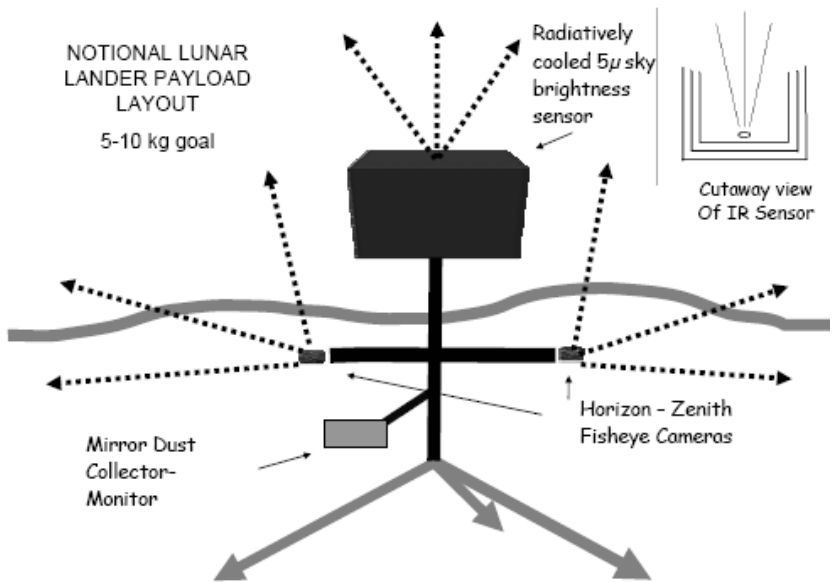

**Fig. 11.** Notional design concept for a micro-Lander to conduct a site test at the pole. Sensors measure dust deposition and any thermal emission from a dust atmosphere above. Cameras view the terrain and sun around the horizon.

**TABLE 1.** Parameters of LLMT designs.

| Diameter (m) | 2 | 20 | 100 |
|---|---|---|---|
| Mirror area (m$^2$) | 3 | 300 | 7600 |
| Mirror density (kg/m$^2$) | 15 | 3.3 | 5-10 |
| Primary mass (tons) | 0.05 | 1 | 50 |
| Total mass (tons) | 0.5 | 3 | 100 |
| Field | 3.1°annulus | 15' | 3' |
| Diffraction limit @ 1μm | 0.1" | 0.01" | 0.002" |
| Pixels @ 2μm (Nyquist) | 18,000 | 45,000 | 45,000 |